# 3D Global climate model of an exo-Venus: a modern Venus-like atmosphere for the nearby super-Earth LP 890-9 c

Diogo Quirino[⊙],[1,2]★ Gabriella Gilli[⊙],[1,3] Lisa Kaltenegger[⊙],[4,5] Thomas Navarro,[6,7] Thomas J. Fauchez,[8,9,10] Martin Turbet,[11] Jérémy Leconte,[12] Sébastien Lebonnois[11] and Francisco González-Galindo[3]

[1]*Instituto de Astrofísica e Ciências do Espaço, Universidade de Lisboa, OAL, Edifício Leste, Tapada da Ajuda, Lisboa PT1349-018, Portugal*
[2]*Faculty of Sciences, University of Lisbon, Campo Grande, Lisboa P-1749-016, Portugal*
[3]*Instituto de Astrofísica de Andalucía (IAA – CSIC), Glorieta de la Astronomía, Granada E-18008, Spain*
[4]*Department of Astronomy, Cornell University, 311 Space Sciences Building, Ithaca, NY 14853, USA*
[5]*Carl Sagan Institute, Cornell University, 302 Space Sciences Building, Ithaca, NY 14853, USA*
[6]*Trottier Space Institute, McGill University, 3550 University Street, Montréal, QC H3A 2A7, Canada*
[7]*Space Science Institute, 4765 Walnut St, Suite B, Boulder, CO 80301, USA*
[8]*NASA Goddard Space Flight Center, Greenbelt, MD 20771, USA*
[9]*Integrated Space Science and Technology Institute, Department of Physics, American University, Washington DC 20016, USA*
[10]*NASA GSFC Sellers Exoplanet Environments Collaboration, 8800 Greenbelt Road, Greenbelt, MD 20771, USA*
[11]*Laboratoire de Météorologie Dynamique, IPSL, CNRS, Sorbonne Université, ENS, PSL Research University, École Polytechnique, Paris F-75005, France*
[12]*Laboratoire d'Astrophysique de Bordeaux, CNRS, Université de Bordeaux, Bât. B18N, Allée Geoffroy Saint-Hilaire, CS 50023, Pessac F-33615, France*

## ABSTRACT

The recently discovered super-Earth LP 890-9 c is an intriguing target for atmospheric studies as it transits a nearby, low-activity late-type M-dwarf star at the inner edge of the Habitable Zone. Its position at the runaway greenhouse limit makes it a natural laboratory to study the climate evolution of hot rocky planets. We present the first 3D-Global Climate Model exo-Venus model for a *modern* Venus-like atmosphere (92 bar surface pressure, realistic composition, and $H_2SO_4$ radiatively-active clouds), applied to the tidally-locked LP 890-9 c to inform observations by *JWST* and future instruments. If LP 890-9 c has developed into a modern exo-Venus, then the modelled temperatures suggest that $H_2SO_4$ clouds are possible even in the substellar region. Like on modern Venus, clouds on LP 890-9 c would create a flat spectrum. The strongest $CO_2$ bands in transmission predicted by our model for LP 890-9 c are about 10 ppm, challenging detection, given *JWST* estimated noise floor. Estimated phase curve amplitudes are 0.9 and 2.4 ppm for *continuum* and $CO_2$ bands, respectively. While pointing out the challenge to characterise modern exo-Venus analogues, these results provide new insights for *JWST* proposals and highlight the influence of clouds in the spectrum of hot rocky exoplanet spectra.

## 1 INTRODUCTION

The recently discovered nearby LP 890-9 planetary system, located at 32 pc, hosts two super-Earths transiting a relatively low-activity M6-type dwarf star. The inner planet was detected by *TESS* in early 2022 and identified as TOI-4306.01. However, the second planet, LP 890-9 c (also identified as SPECULOOS-2 c), was discovered with the follow-up ground-based SPECULOOS search by Delrez et al. (2022) and is a rocky planet on the inner edge of the Habitable Zone (HZ), the temperate zone around a star, where water could remain liquid on the surface on an Earth-like planet (see Kasting, Whitmire & Reynolds 1993; Kopparapu et al. 2013, 2017; Kaltenegger 2017). This location makes LP 890-9 c an incredible natural laboratory to understand the comparative climate evolution between Earth and Venus. Furthermore, it raises the question of whether the planet supports an Earth-like or a Venus-like climate (Kaltenegger et al. 2022, see). LP 890-9 c is an intriguing target to study because it can lead to insights into rocky planet evolution for increased incident irradiation.

The two planets have similar sizes (planet b: 1.32 $R_\oplus$; and planet c: 1.37 $R_\oplus$) and orbit the host star in 2.73 and 8.46 d, respectively (Delrez et al. 2022). The inner planet has an irradiation about double that of Venus (4.09 ± 0.12 $S_\oplus$). LP 890-9 b orbits very close to the star inwards of the runaway greenhouse limit but outside an inner limit of estimated atmospheric erosion in the 'Venus Zone' (Kane, Kopparapu & Domagal-Goldman 2014; see Fig. 1). The outer planet LP 890-9 c receives a stellar flux a bit lower than modern Earth (0.906 ± 0.026 $S_\oplus$), placing it inside the conservative HZ (Kopparapu et al. 2014, 2017). However, because it orbits an M star, this stellar irradiation puts it very close to the runaway greenhouse limit (e.g. Kasting et al. 1993; Kopparapu et al. 2013; Kaltenegger et al. 2022). The discovery paper (Delrez et al. 2022) points out that LP 890-9 c is among the best targets for the atmospheric characterization of rocky planets in the HZ for *JWST* and future observatories (Extremely Large Telescopes, ELTs) depending on its evolution from a hot Earth to a modern Venus (Kaltenegger et al. 2022), in line with TRAPPIST-1 (aka SPECULOOS-1) planetary system (Gillon et al. 2016, 2017; Agol et al. 2021).

★ E-mail: dfquirino@fc.ul.pt

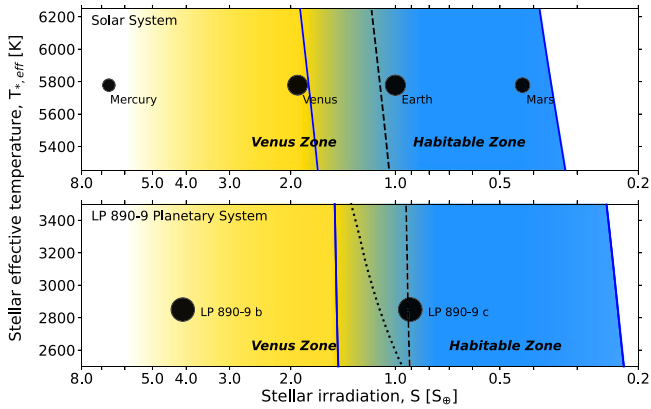

**Figure 1.** Stellar irradiation comparison for the four inner planets of the Solar System (upper panel) and the LP 890-9 planetary system (lower panel), with scaled planetary radii. The Venus Zone (yellow shade) (Kane et al. 2014) and HZ (blue shade) transition gradually due to the complex climate feedback that may push planets into Earth-like or Venus-like states within the orbital range. HZ limit (solid blue) shown: inner empirical 'Recent Venus'; and outer 'Maximum Greenhouse' limit. A 'Runaway Greenhouse' flux limit is also indicated following 1D (black dashed) (Kopparapu et al. 2014) and 3D (black dotted) climate modelling (Kopparapu et al. 2017).

Atmosphere evolution in M-dwarf planets can be heavily impacted (Ramirez & Kaltenegger 2014; Luger & Barnes 2015) by the long, extremely active, superluminous pre-main-sequence phase, which can last ∼600 Myr for an M6V star (Baraffe et al. 2002), and the life-long stellar activity (e.g. Scalo et al. 2007; Tarter et al. 2007), whose flares, coronal mass ejections (CMEs), high-energy radiation, and charged particles can drive thermal and non-thermal escape processes resulting in atmosphere erosion or extensive modification (Lammer et al. 2007; Scalo et al. 2007; Kay, Opher & Kornbleuth 2016; Airapetian et al. 2017; Tilley et al. 2019). In addition, solar events like CMEs are known to lead to atmosphere loss and long-term evolution (e.g. Jakosky et al. 2015, 2018; Mayyasi et al. 2018). Thus, an atmosphere on LP 890-9 c has likely evolved from its primordial composition.

Carbon dioxide-dominated atmospheres are a possible scenario for these secondary atmospheres (Turbet et al. 2020a; Kaltenegger et al. 2022). Due to carbon dioxide's heavy molecular weight and very efficient IR cooling, $CO_2$-dominated atmospheres have extremely cold thermospheres and a less expanded upper atmosphere, leading to larger resilience to thermal and non-thermal escape processes (Lammer et al. 2008; Tian 2009; Cohen et al. 2015; Johnstone et al. 2018; Gronoff et al. 2020; Turbet et al. 2020a) than atmospheres of other most common chemical species (e.g. $N_2$, $O_2$, or $H_2$), which are expected to escape on shorter time-scales (Kulikov et al. 2006; Lammer, Kulikov & Lichtenegger 2006; Lammer et al. 2007; Tian 2009; Airapetian et al. 2017; Garcia-Sage et al. 2017; Tilley et al. 2019). Dense $CO_2$ atmospheres of super-Earths in the HZ might also endure thermal escape during the high-activity levels of M-dwarfs (Tian 2009). However, even for Venus-sized, non-magnetized, close-in exoplanets orbiting M-dwarfs, the lower limit of atmospheric mass-loss rate from stellar wind suggests that a Venus-like atmosphere could be sustainable for billions of years around such stars (Cohen et al. 2015). Given the extensive atmospheric evolution, $CO_2$ atmospheric build-up is a likely climate evolution scenario for terrestrial planets orbiting M-dwarf stars (Turbet et al. 2020a).

Here, we use a 3D Global Climate Model (GCM), the Generic-PCM, to present 3D simulations for a tidally-locked, modern exo-Venus atmosphere for LP 890-9 c. It is the first time that a 3D climate simulation of a ground-to-the-upper atmosphere for a modern 'exo-Venus' is performed using the same atmospheric parameters as on present-day Venus. The atmosphere has a 92 bar surface pressure, realistic composition, and a global cover of sulphuric acid aerosols.

Due to its instellation and proximity to the runaway greenhouse limit, LP 890-9 c provides an exciting opportunity to test the limits of the HZ's inner edge. In addition, observation of LP 890-9 c can also provide insight into where the climate divergence between Earth and Venus originates from (see also Turbet et al. 2021; Kaltenegger et al. 2022) and explore surface habitability on rocky exoplanets.

We introduce the model's details and the parameters selection rationale in Section 2. Then, in the results and discussion (Section 3), we present the predicted thermal structure and large-scale circulation as well as synthetic observables (i.e. thermal phase curves and transmission spectra) to assess the potential for future observations. Finally, we summarize the main points of our work in Section 4.

## 2 METHODS

We model the atmosphere of LP 890-9 c with the Generic-PCM (Planetary Climate Model), formerly known as the LMD Generic GCM, specifically developed for exoplanet and paleoclimate studies at the Laboratoire de Météorologie Dynamique (Institut Pierre – Simon Laplace, Paris, France). The Generic-PCM historically derives from the LMDz Earth GCM (Hourdin et al. 2006) and LMD Mars GCM (Forget et al. 1999), which solves the primitive equations of geophysical fluid dynamics using a finite difference 3D dynamical core (Charnay, Meadows & Leconte 2015). The model uses an up-to-date generalized radiative transfer routine (Wordsworth, Forget & Eymet 2010) and correlated-k tables from kspectrum (Eymet, Coustet & Piaud 2016), an open-source code for high-resolution molecular absorption spectral production, and has been used to simulate different atmospheres and climates across a wide irradiation range, both in the Solar System and in exoplanets (e.g. Wordsworth et al. 2011; Leconte et al. 2013; Wordsworth et al. 2013; Charnay et al. 2014; Turbet et al. 2016, 2018, 2020b).

We assumed in our model that LP 890-9 c has a circular orbit with no obliquity (no seasons) and is tidally-locked, based on the estimated age of the system (7.2 Gyr) (Delrez et al. 2022), which is longer than the time required to tidally lock the planet ($10^6$ yr) and for orbital circularization ($\sim 10^9$–$\sim 10^7$ yr, depending on the dissipation factor). The discovery paper (Delrez et al. 2022) estimates an equilibrium eccentricity of $\sim 1.7 \times 10^{-5}$, resulting from planet–planet interactions and tidal damping (Turbet et al. 2018), with orbits having the time to evolve to very low eccentricities as a consequence of tidal interactions. Assuming a synchronous tidally-locked configuration (rotation and orbital periods are equal) led to a sidereal day period of 730 700 s (rotation rate, $8.6 \times 10^{-6}$ rad s$^{-1}$). The remaining planetary parameters were taken directly from the discovery paper (Delrez et al. 2022) and are listed in Table 1. The horizontal resolution is 64 × 48 in longitude and latitude (i.e. 5.626° longitude by 3.75° latitude), ∼855 km per 570 km at the equator of LP 890-9 c. We assumed a rocky continental flat surface with thermal

**Table 1.** Planetary parameters of LP 890-9 c used in the Generic-PCM.

| Planetary parameter | LP 890-9 c |
| --- | --- |
| Orbital period, $T_P$ (d) | 8.46 |
| Semimajor axis, $a$ (au) | 0.03984 |
| Instellation, $S$ (W m$^{-2}$) | 1233.5 |
| Radius, $R$ (km) | 8709 |
| Gravitational acceleration, $g$ (m s$^{-2}$) | 13.35 |
| Solid-body rotation period, $P$ (d) | 8.46 |

*Note.* Eccentricity and obliquity are set to 0.

inertia of 2000 J m$^{-2}$ K$^{-1}$ s$^{-1/2}$, following Leconte et al. (2013). We do not consider internal heat flux in this study. We use the stellar spectrum from Delrez et al. (2022; their fig. 6), ranging from 0.1 to 24.1 µm, as an input for our simulations.

In this study, we simulate a CO$_2$-dominated (i.e. 96.5 per cent) modern Venus-like atmosphere for LP 890-9 c, with a surface pressure of 92 bar and a homogeneous, global cloud cover of sulphuric acid aerosols. We have included four minor components: SO$_2$ (0.07–180 ppm), OCS (0.01–51 ppm), H$_2$O (1.04–31.1 ppm), and CO (10–43.9 ppm) with abundances similar to modern Venus for both the radiative transfer suite and Rayleigh diffusion calculations. We assume a specific heat capacity, $c_p$, uniformly fixed at 1061.4 J kg$^{-1}$ K$^{-1}$.

The model contains 50 atmospheric layers from the ground to $p \sim$ 5 Pa, with hybrid $\sigma - p$ coordinates for vertical discretization. The simulation was initialized using a start file from an atmosphere in superrotation obtained from the IPSL Venus GCM (Lebonnois, Sugimoto & Gilli 2016). This initial start file was modified to create a new start file, taking into account the LP 890-9 c planetary parameters listed in Table 1. The simulation runs until a quasi-radiative balance is reached ($< 1$ W m$^{-2}$), following the methodology in Turbet et al. (2021). The IPSL Venus GCM succeeds in reproducing the atmosphere's superrotation and actual Venus's global temperature field. Given the nature of our simulations and the assumption of a Venus-like atmosphere, we consider this initialization a reasonable choice. We included a prescribed globally-uniform cloud cover between 1 and 0.037 bar following observations of Venus Express/ESA (Haus, Kappel & Arnold 2013, 2015) and previous model development done for the IPSL Venus GCM (Garate-Lopez & Lebonnois 2018). The aerosol optical depth is computed at each grid box using four sulphuric acid particle modes (1, 2, 2p, and 3) and the 'unknown' UV absorber (see Haus et al. 2013, 2015). The resulting cloud mixing ratio is held fixed throughout the simulation, and there are no tracers advected by the dynamics.

## 3 RESULTS AND DISCUSSION

Fig. 2 depicts the mean thermal structure of our LP 890-9 c modern Venus model: global (blue); substellar region (orange) centred on the substellar point (longitude 0°) and spatially-averaged between 15° N/S and 17° W/E; nightside hemisphere (dark blue); and nightside equatorial (red) region (averaged between 15° N/S). The Venus International Reference Atmosphere (VIRA) (black) profile (Keating et al. 1985; Seiff et al. 1985) is shown for reference.

Although M-dwarf stars emit mostly in the IR and are more effective in heating the surface of an Earth-like planet (e.g. Kasting et al. (1993)), LP 890-9 c deep atmosphere below 0.1 bar is globally colder than Venus', with a surface temperature of 546 K (versus 735 K on Venus). Interpreting this colder surface temperature is not straightforward and requires a detailed analysis. Several factors can contribute to cooling down the temperature in the deep atmosphere of the planet: the lower bolometric instellation than on Venus, the assumption of a fixed $c_p$ (compared with the temperature-dependent $c_p$ on Venus), the chosen aerosol parametrization, and the large sensitivity of troposphere to opacity sources.

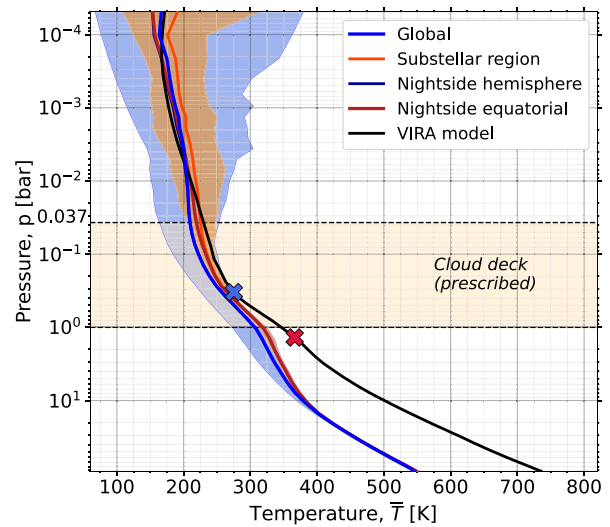

**Figure 2.** Temperature-pressure profiles of the modern Venus LP 890-9 c model averaged over 10 orbits for four different regions (see text for details): global (blue); substellar region (orange); nightside hemisphere (dark blue); and nightside equatorial (red). The shaded areas (global and substellar region) depict the temperature value range at each pressure level. The VIRA temperature profile is shown (black) for comparison. The prescribed global cloud deck (pale orange shade) extends between 1 and 0.037 bar (dashed black lines). The red (blue) cross represents the condensation (freezing) point of sulphuric acid for Venus-like conditions.

The lapse rate in our model is $-11.8$ K km$^{-1}$ from the surface to $\sim$10 bar, decreasing to $-6.6$ K km$^{-1}$ for the region between $\sim$10 bar and the cloud base ($p \sim 1$ bar), and increasing to $-10.0$ K km$^{-1}$ between the cloud base and $\sim$0.1 bar. Above the cloud top (p $\sim$ 37 mbar), the global profile is nearly isothermal and similar to VIRA, particularly above $\sim$5 mbar. The cold upper atmosphere reflects the efficient CO$_2$ cooling at 15-µm in our 3D model (Schubert et al. 1980; Gilli et al. 2017).

All temperature profiles in Fig. 2 are isothermal above the cloud top, with the larger stellar energy absorption leading to a warmer (on average, $\sim$13 K) substellar region. This region remains significantly warmer than the global average through the atmosphere and down to 10 bar, where the temperature differences become smaller than 5 K. On the contrary, the nightside equatorial profile is similar to the substellar region (maximum difference: 5 K) below the cloud top and approaching the global profile above this level with temperature differences smaller than 5 K above $\sim$4 mbar. The similar temperatures suggest warm air advection from the dayside to the nightside in the low latitudes and across a wide pressure range.

The thermal structure of LP 890-9 c is colder than Venus at the sulphuric acid condensation and freezing points pressure levels, shown as red and blue crosses in Fig. 2, respectively. A colder atmosphere suggests that sulphuric acid cloud formation and stability at the prescribed cloud base might be possible globally under simulated conditions. Furthermore, this could indicate that the cloud base could reach deeper into the atmosphere. However, several studies (e.g. France et al. 2013; Gao et al. 2015; Jordan et al. 2021) suggest changes in the composition and opacity sources for terrestrial exoplanets around cooler and highly active M-dwarf stars, compared with the present-day Venus atmosphere. For instance, Jordan et al. (2021) showed that sulphuric acid molecule production decreases with decreasing stellar effective temperature and that resulting trace sulphur gases can be prime observational indicators, potentially distinguishing between Venus and Earth-like climates around M-dwarf stars. Therefore, the impact of these compositional changes on the cloud deck parameters (e.g. altitude range) should be considered in a future analysis.

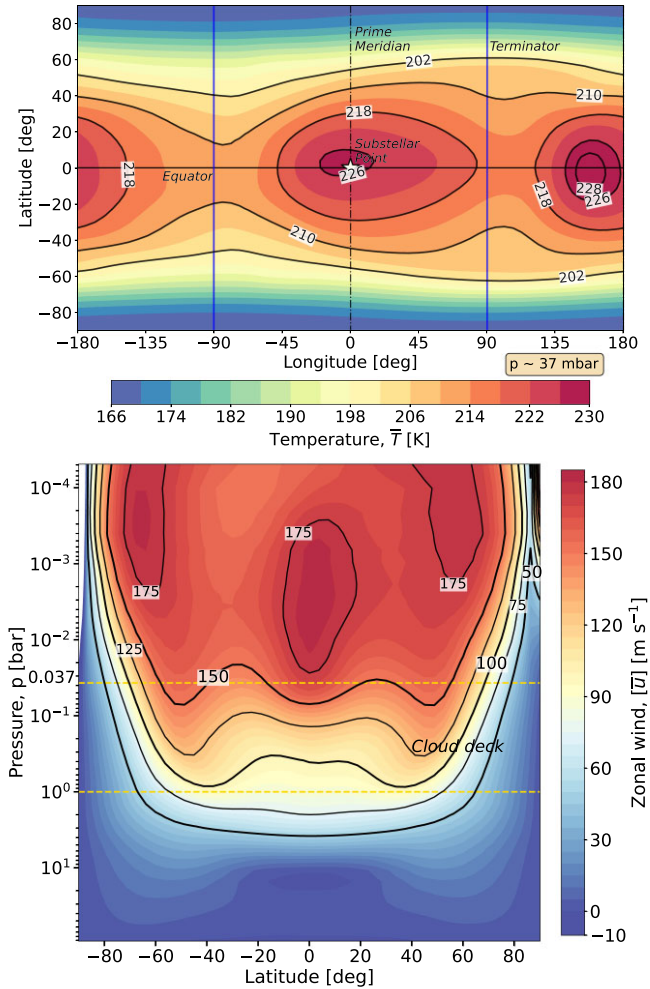

**Figure 3.** (top) Time-averaged cloud top ($p \sim 37$ mbar) temperature field for 10 orbits of LP 890-9 c. The white star marks the substellar point (latitude 0° and longitude 0°). (bottom) Zonally and time-averaged zonal wind field for 10 orbits of LP 890-9 c. The solid black lines represent thick (thin) isotachs with an interval of 50 m s$^{-1}$ (25 m s$^{-1}$).

The cloud top temperature (Fig. 3), close to the infrared photosphere, influences the thermal phase curves (Showman & Guillot 2002). Fig. 3 shows two hot spots, a substellar and a slightly warmer nightside hotspot, located west of the antistellar point (∼170° E). A substellar hotspot was also generated in other 3D simulations of a tidally-locked planet with an orbital period of 10 d (Pierrehumbert & Hammond 2019). The presence of cold Rossby cyclones in the 90° W terminator region and a chevron-like pattern in the dayside hemisphere, together with the nightside hotspot, indicate an equatorial zonal superrotation jet – a classical modelling result for tidally-locked planets (Showman & Polvani 2011; Heng & Showman 2015; Pierrehumbert & Hammond 2019).

The zonally and time-averaged zonal wind field (see Fig. 3) depicts an atmosphere under a superrotation regime, with three eastward superrotation jets: an equatorial and two high-latitudes jets centred at ∼60° N/S. The superrotation equatorial jet is a classical modelling result for tidally-locked planets (Showman & Polvani 2011), while the high-latitudes jets are a consequence of a slowly rotating atmosphere (Heng & Showman 2015). All jets have cores above the cloud top with maximum speeds of ∼180 m s$^{-1}$. The jets' cores are located at different pressure levels, with the equatorial jet between 30 and 0.3 mbar, deeper than the high-latitudes jets, which extend above 2 mbar. The jets are present at the cloud top level across all longitudes, with the largest speeds at the terminators, where the equatorial jet reaches 180 m s$^{-1}$ east of the 90° W terminator.

The dynamics obtained in our simulation with a modern Venus-like atmosphere substantially differ from previous simulations of tidally locked terrestrial planets with Earth-like atmospheres (surface pressure of 1 bar) but with similar stellar irradiation and rotation rate values to this work. These studies consistently found a single equatorial superrotation jet located at ∼500 mbar with speeds on the order of dozens of m s$^{-1}$ (Penn & Vallis 2018; Hammond, Tsai & Pierrehumbert 2020; Hammond & Lewis 2021; Sergeev et al. 2022; Wang & Yang 2022). However, at least two distinct dynamical regimes can occur in these worlds (e.g. Noda et al. 2017; Carone et al. 2018; Sergeev et al. 2022). For example, Sergeev et al. (2022) shows that a single equatorial jet or two mid-latitude jets regime is possible. This climate bistability highly depends on the model's initial conditions and the convection and cloud radiative parametrization schemes. In addition, Eager-Nash et al. (2020), in their simulation of TRAPPIST-1 e with an Earth-like atmosphere and water clouds, also obtained two superrotating jets at 30° N/S. In these studies, the impact of convection and cloud radiative schemes is significant for the resulting dynamics. They will define the relative absorption of stellar energy at each atmospheric level and on the surface.

### 3.1 Synthetic transmission and emission

Using our GCM outputs, aerosols and chemical species abundances and the global temperature profile, we also explored the potential to detect molecular species on LP 890-9 c during transits observed by *JWST*. We simulate the transmission spectra with the Planetary Spectrum Generator (Villanueva et al. (2018)) assuming Venus-like clouds shown in Fig. 4 (top panel) for NIRSpec Prism on *JWST*. The simulations show that high-altitude sulphuric acid clouds would obscure most spectral features, except for the 2.8 and 4.3-$\mu$m $CO_2$ in a Venus-like atmosphere. The largest $CO_2$ feature peaks at 10 ppm in our model, barely reaching NIRSpec Prism's estimated noise floor. Following the method described in Fauchez et al. (2022) (see equations 2 and 3), we estimated a minimum number of 1039 and 2841 transits needed for a $CO_2$ 5$\sigma$ detection with NIRSpec Prism and NIRISS-SOSS (not shown), respectively, for a modern-Venus atmosphere for LP 890-9 c, using a factor of three for the out-of-transit time (Fauchez et al. 2022). To show clearly the strong impact of the clouds and Mie scattering on the transmission spectra, we also artificially removed clouds from the terminator atmospheric profiles (black dashed curve in Fig. 4).

The bottom panel of Fig. 4 provides the expected emission phase curves for a modern Venus atmosphere for LP 890-9 c using OLR fluxes computed by the Generic-PCM over 32 spectral bands (between 3.51 and 22.73 $\mu$m). These radiative fluxes were integrated over the Earth-facing hemisphere across several orbital phases. The emission phase curves are calculated as the ratio between the planet's emitted flux and the stellar flux in the same spectral band, using the specifications in Von Paris et al. (2016). The emission phase curves are presented for two different spectral bands: the *continuum* (11.43–13.89 μm) and the 15-μm carbon dioxide absorption band (13.89–16.21 μm). Overall, the shape of the phase curve is similar between the two bands. The largest contrast values are between 5 and 7 ppm for the *continuum* and $CO_2$ absorption bands, respectively. The largest amplitude modulation occurs between the secondary eclipse and the transit for both phase curves. However, the amplitude is 2.4 ppm for the $CO_2$ absorption band and less than 1 ppm for the *continuum* (double-head arrow values in Fig. 4), making the detection challenging with current instruments.

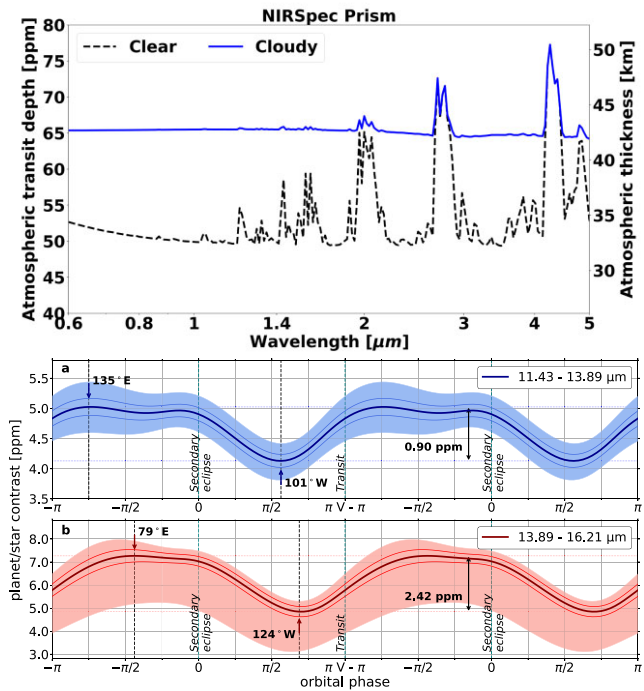

**Figure 4.** (top) Simulated transmission spectra of LP 890-9 c based on Generic-PCM outputs for Venus-like clouds (blue) and clear-sky conditions (black) for `NIRSpec Prism` (from 0.6 to 5 μm). (bottom) Emission phase curves time-averaged over 10 orbits for two bands: (a) *continuum*: 11.43–13.89 μm; and (b) 15-μm carbon dioxide absorption: 13.89–16.21 μm. The time-averaged (1σ) phase curve is shown as solid curves, while the shaded region represents the range of possible planet-to-star contrast values. The dashed black lines and arrows with the corresponding sub-observer longitude indicate the orbital phase of time-averaged maximum and minimum contrast. A black double-head arrow depicts the phase curve amplitude, whose limits are marked by dashed lines. The simulations are displayed over two orbits with the secondary eclipse and transit occurring at orbital phase 0 and $-\pi/\pi$ (vertical dashed green lines), respectively.

Though we have used a fixed Venus-like cloud mixing ratio in this study, we can speculate about the impact of varying this parameter in the emission phase curves. For instance, a 10-fold increase implies that more stellar radiation is absorbed in the upper levels of the cloud deck, possibly leading to a stronger jet in the upper atmosphere as more energy becomes available in this atmospheric region. A stronger zonal advection of warm air masses at the cloud top level would lead to a larger eastward offset of the hotspot from the substellar point. In this situation, the phase curve would peak closer to longitude 180°, i.e. further away from the secondary eclipse. Conversely, a 10-fold decrease allows deeper radiation penetration within the cloud deck. In this scenario, a weaker jet would form, leading to a smaller hotspot offset with phase curve peak emission occurring closer to the substellar longitudes/secondary eclipse. Naturally, the precise assessment of the physical consequences of adopting any of these scenarios would require running the model from the start until an acceptable radiative balance can be achieved.

## 4 CONCLUSIONS

The position of the recently discovered super-Earth LP 890-9 c at the runaway greenhouse limit makes it a natural laboratory to study the climate evolution of hot rocky planets. Assuming a tidally-locked planet that has evolved to a *modern* Venus-like atmosphere, our 3D climate simulations with the Generic-PCM show an effective day–night heat redistribution by three superrotation jets: one equatorial and two high-latitudes. If LP 890-9 c has developed into a modern exo-Venus, then the modelled temperatures suggest that $H_2SO_4$ clouds are possible even in the substellar region, creating a flat transmission spectrum. The strongest $CO_2$ bands in transmission predicted by our model are about 10 ppm, challenging given *JWST* estimated noise floor. Estimated phase curve amplitudes are 0.9 and 2.4 ppm for *continuum* and $CO_2$ bands, respectively. The results presented in this paper are based on the hypothesis that LP 890-9 c has evolved into a modern Venus-like atmosphere and is in synchronous rotation, an assumption that can be tested with future observations. The atmospheric composition of such small and rocky planets could be only accessible to the Large Interferometer for Exoplanets (LIFE) (Konrad et al. 2022). LP 890-9 c is a rare laboratory for atmosphere characterization and a benchmark to understand possible paths of terrestrial climate evolution under higher stellar irradiation and the inner limits of the HZ.

## ACKNOWLEDGEMENTS

The authors thank the Generic-PCM team for the teamwork development and improvement of the model, in particular E. Millour for his invaluable support. DQ and GG acknowledge support from the Portuguese FCT (project P-TUGA Ref. PTDC/FIS-AST/29942/2017 through national funds and by FEDER through COMPETE 2020, Ref. POCI-01-0145 FEDER-007672). GG and FGG acknowledge support from the Spanish MCIU, the AEI, and EC FEDER (project RTI2018100920JI00) and financial support from the grant CEX2021-001131-S funded by MCIN/AEI/ 10.13039/501100011033. GG acknowledges financial support from Junta de Andalucia through the program EMERGIA 2021 (EMC21_00249). LK acknowledges support from the Brinson Foundation ('Search for Life in the Universe') and the Carl Sagan Institute at Cornell Univ. MT acknowledges support from BELSPO BRAIN (project B2/212/PI/PORTAL) and the Tremplin 2022 program (Faculty of Science and Engineering, Sorbonne Univ.). TN acknowledges support from the Trottier Space Institute. TJF acknowledges support from the GSFC Sellers Exoplanet Environments Collaboration (SEEC), which is funded in part by the NASA Planetary Science Divisions Internal Scientist Funding Model.

## DATA AVAILABILITY

The outputs of the model data: temperature profiles, emission phase curves and transmission spectra calculated for `NIRSpec Prism`, as well as larger versions of the synthetic observables figures, are available through the following Zenodo link: https://doi.org/10.5281/zenodo.7671735

This paper has been typeset from a T<sub>E</sub>X/L<sup>A</sup>T<sub>E</sub>X file prepared by the author.